# A New Viewpoint to the Discrete Approximation: Discrete Yang-Fourier Transforms of Discrete-Time Fractal Signal

Xiao-Jun Yang

*Department of Mathematics and Mechanics, China University of Mining & Technology, Xuzhou, Jiangsu, P. R. China.*

Email:dyangxiaojun@163.com

**Abstract**

It is suggest that a new fractal model for the Yang-Fourier transforms of discrete approximation based on local fractional calculus and the Discrete Yang-Fourier transforms are investigated in detail.

***Key words:*** *local fractional calculus, fractal, Yang Fourier transforms, discrete approximation, discrete Yang-Fourier transforms*

MSC2010: 28A80, 26A99, 26A15, 41A35

## 1 Introduction

Fractional Fourier transform becomes a hot topic in both mathematics and engineering. There are many definitions of fractional Fourier transforms [1-5]. Hereby we write down the Yang-Fourier transforms [3-5]

$$F_\alpha\{f(x)\} = f_\omega^{F,\alpha}(\omega) := \frac{1}{\Gamma(1+\alpha)} \int_{-\infty}^{\infty} E_\alpha(-i^\alpha \omega^\alpha x^\alpha) f(x)(dx)^\alpha, \quad (1.1)$$

and its inverse, denoted by [3,4]

$$f(x) = F_\alpha^{-1}\left(f_\omega^{F,\alpha}(\omega)\right) := \frac{1}{(2\pi)^\alpha} \int_{-\infty}^{\infty} E_\alpha(i^\alpha \omega^\alpha x^\alpha) f_\omega^{F,\alpha}(\omega)(d\omega)^\alpha, \quad (1.2)$$

where local fractional integral of $f(t)$ is denoted by[3-8]

$$\begin{aligned}
{}_aI_b^{(\alpha)} f(x) &= \frac{1}{\Gamma(1+\alpha)} \int_a^b f(t)(dt)^\alpha \\
&= \frac{1}{\Gamma(1+\alpha)} \lim_{\Delta t \to 0} \sum_{j=0}^{j=N-1} f(t_j)(\Delta t_j)^\alpha
\end{aligned} \quad (1.3)$$

with $\Delta t_j = t_{j+1} - t_j$ and $\Delta t = \max\{\Delta t_1, \Delta t_2, \Delta t_j, ...\}$, where for $j = 0,...,N-1$, $[t_j, t_{j+1}]$ is a partition of the interval $[a,b]$ and $t_0 = a, t_N = b$. Here, for $|x - x_0| < \delta$ with $\delta > 0$, there exists any $x$ such that

$$|f(x) - f(x_0)| < \varepsilon^\alpha. \quad (1.4)$$

Now $f(x)$ is called local fractional continuous at $x = x_0$ and we have [5]

$$\lim_{x \to x_0} f(x) = f(x_0). \quad (1.5)$$

Suppose that $\{f_0, f_1, \cdots, f_{N-1}\}$ is an $N_{th}$ order regular sampling with spacing $\Delta x$ some piecewise local fractional continuous function over mammal window $[0, L]$. In the present paper, our arms are to get some assurance that local fractional integral of $f$ can be reasonably approximated by the corresponding integration of $\tilde{f}$ and we will get the discrete Yang-Fourier transforms.

## 2 A fractal model for the Yang-Fourier transforms of discrete approximation

Now we determine from our data,

$$\frac{1}{\Gamma(1+\alpha)}\int_{-\frac{1}{2}\Delta t}^{\frac{2N-1}{2}\Delta t}\tilde{f}(t)\phi(t)(dt)^{\alpha} \approx \frac{1}{\Gamma(1+\alpha)}\int_{-\frac{1}{2}\Delta t}^{\frac{2N-1}{2}\Delta t} f(t)\phi(t)(dt)^{\alpha} \quad (2.1)$$

for any local fractional continuous function on the natural widow. This sampling can be used to complete a corresponding sum approximation for the integration,

$$\frac{1}{\Gamma(1+\alpha)}\int_{-\frac{1}{2}\Delta t}^{\frac{2N-1}{2}\Delta t} f(t)\phi(t)(dt)^{\alpha}$$
$$\approx \frac{1}{\Gamma(1+\alpha)}\sum_{k=0}^{N-1} f(k\Delta t)\phi(k\Delta t)(\Delta t)^{\alpha} \quad (2.2)$$
$$= \frac{1}{\Gamma(1+\alpha)}\sum_{k=0}^{N-1} f_k \phi(k\Delta t)(\Delta t)^{\alpha}.$$

Notice, however, that

$$\frac{1}{\Gamma(1+\alpha)}\sum_{k=0}^{N-1} f_k \phi(k\Delta t)(\Delta t)^{\alpha}$$
$$= \frac{1}{\Gamma(1+\alpha)}\sum_{k=0}^{N-1} f_k \left[\frac{1}{\Gamma(1+\alpha)}\int_{-\frac{1}{2}\Delta t}^{\frac{2N-1}{2}\Delta t}\phi(t)\delta_{k\Delta t}(t)(dt)^{\alpha}\right](\Delta t)^{\alpha}$$
$$= \frac{1}{\Gamma(1+\alpha)}\int_{-\frac{1}{2}\Delta t}^{\frac{2N-1}{2}\Delta t}\left[\frac{1}{\Gamma(1+\alpha)}\sum_{k=0}^{N-1} f_k \delta_{k\Delta t}(t)(\Delta t)^{\alpha}\right]\phi(t)(dt)^{\alpha} \quad (2.3)$$

where

$$\frac{1}{\Gamma(1+\alpha)}\int_{-\frac{1}{2}\Delta t}^{\frac{2N-1}{2}\Delta t}\phi(t)\delta_{k\Delta t}(t)(dt)^{\alpha} = \phi(k\Delta t), \text{ for } k=0,1,\cdots,N-1$$

So,

$$\frac{1}{\Gamma(1+\alpha)}\int_{-\frac{1}{2}\Delta t}^{\frac{2N-1}{2}\Delta t} f(t)\phi(t)(dt)^{\alpha}$$
$$= \frac{1}{\Gamma(1+\alpha)}\int_{-\frac{1}{2}\Delta t}^{\frac{2N-1}{2}\Delta t}\left[\frac{1}{\Gamma(1+\alpha)}\sum_{k=0}^{N-1} f_k \delta_{k\Delta t}(t)(\Delta t)^{\alpha}\right]\phi(t)(dt)^{\alpha} \quad (2.4)$$

Suggesting that, with the natural window, we use

$$\tilde{f}(t) = \frac{1}{\Gamma(1+\alpha)}\sum_{k=0}^{N-1}\tilde{f}_k \delta_{k\Delta t}(t), \quad (2.5)$$

where $\tilde{f}_k = f_k(\Delta t)^{\alpha}$ for $k=0,1,\cdots,N-1$.

Now there are two natural choices: Either $\tilde{f}$ define to be 0 outside the nature window, or define $\tilde{f}$ to be periodic with period $T$ equalling the length of the natural window,

$$T = N\Delta t. \quad (2.7)$$

Combing with our definition of $\tilde{f}$ on the natural window, the first choice would be give

$$\tilde{f}(t) = \frac{1}{\Gamma(1+\alpha)} \sum_{k=0}^{N-1} \tilde{f}_k \delta_{k\Delta t}(t), \qquad (2.8)$$

while the second choice would be give

$$\tilde{f}(t) = \frac{1}{\Gamma(1+\alpha)} \sum_{k=-\infty}^{\infty} \tilde{f}_k \delta_{k\Delta t}(t) \qquad (2.9)$$

with $\tilde{f}_{k+N} = \tilde{f}_k$.

Clearly, the latter is the more clear choice. That is to say, suppose that $\{f_0, f_1, \cdots, f_{N-1}\}$ is the $N_{th}$ order regular sampling with spacing $\Delta t$ of some function $f$. The corresponding discrete approximation of $f$ is the periodic, regular array

$$\tilde{f}(t) = \frac{1}{\Gamma(1+\alpha)} \sum_{k=-\infty}^{\infty} \tilde{f}_k \delta_{k\Delta t}(t) \qquad (2.10)$$

with spacing $\Delta t$ index period $N$, and its coefficients

$$\tilde{f}_k = \begin{cases} f_k (\Delta x)^\alpha, & \text{if } k = 0, 1, \cdots, N-1. \\ f_{k+N}, & \text{in general}. \end{cases} \qquad (2.11)$$

From the Yang-Fourier transform theory, we then know

$$F_\alpha\{f(x)\} = f_\omega^{F,\alpha}(\omega)$$

is a local fractional continuous and is given by

$$f_\omega^{F,\alpha}(\omega)$$
$$= \frac{1}{\Gamma(1+\alpha)} \int_{-\infty}^{\infty} f(t) E_\alpha(-i^\alpha \omega^\alpha t^\alpha)(dt)^\alpha$$
$$= \frac{1}{\Gamma(1+\alpha)} \int_{-\frac{1}{2}\Delta t}^{\frac{2N-1}{2}\Delta t} f(t) E_\alpha(-i^\alpha \omega^\alpha t^\alpha)(dt)^\alpha$$
$$\approx \frac{1}{\Gamma(1+\alpha)} \int_{-\frac{1}{2}\Delta t}^{\frac{2N-1}{2}\Delta t} \tilde{f}(t) E_\alpha(-i^\alpha \omega^\alpha t^\alpha)(dt)^\alpha$$
$$= \frac{1}{\Gamma(1+\alpha)} \int_{-\frac{1}{2}\Delta t}^{\frac{2N-1}{2}\Delta t} \left( \frac{1}{\Gamma(1+\alpha)} \sum_{k=0}^{N-1} f_k \delta_{k\Delta t}(t)(\Delta t)^\alpha \right) E_\alpha(-i^\alpha \omega^\alpha t^\alpha)(dt)^\alpha$$
$$= \frac{1}{\Gamma(1+\alpha)} \sum_{k=0}^{N-1} f_k (\Delta t)^\alpha \left( \frac{1}{\Gamma(1+\alpha)} \int_{-\frac{1}{2}\Delta t}^{\frac{2N-1}{2}\Delta t} \delta_{k\Delta t}(t) E_\alpha(-i^\alpha \omega^\alpha t^\alpha)(dt)^\alpha \right)$$
$$= \frac{1}{\Gamma(1+\alpha)} \sum_{k=0}^{N-1} f_k (\Delta t)^\alpha E_\alpha(-i^\alpha \omega^\alpha k^\alpha (\Delta t)^\alpha) \qquad (2.12)$$

So, approximation of the formula

$$\frac{1}{\Gamma(1+\alpha)} \int_{-\infty}^{\infty} f(t) E_\alpha(-i^\alpha \omega^\alpha t^\alpha)(dt)^\alpha$$

reduces to

$$f_\omega^{F,\alpha}(\omega) \approx \frac{1}{\Gamma(1+\alpha)} \sum_{k=0}^{N-1} f_k (\Delta t)^\alpha E_\alpha(-i^\alpha \omega^\alpha k^\alpha (\Delta t)^\alpha). \qquad (2.13)$$

with $T = N\Delta t$.

Taking $\omega = n\Delta\omega$ and $\dfrac{2\pi}{T} = \Delta\omega$ in (2.13) implies that

$$\phi(n)$$
$$= f_\omega^{F,\alpha}(\omega)$$
$$\approx \frac{1}{\Gamma(1+\alpha)} \sum_{k=0}^{N-1} f_k (\Delta t)^\alpha E_\alpha\left(-i^\alpha \omega^\alpha k^\alpha (\Delta t)^\alpha\right) \quad (2.14)$$
$$= \frac{1}{\Gamma(1+\alpha)} \frac{T^\alpha}{N^\alpha} \sum_{k=0}^{N-1} f_k E_\alpha\left(-i^\alpha (2\pi)^\alpha n^\alpha k^\alpha / N^\alpha\right)$$
$$= \frac{1}{\Gamma(1+\alpha)} \frac{T^\alpha}{N^\alpha} \sum_{k=0}^{N-1} \varphi(k) E_\alpha\left(-i^\alpha (2\pi)^\alpha n^\alpha k^\alpha / N^\alpha\right)$$

In the same manner, if
$$f(t) = \frac{1}{(2\pi)^\alpha} \int_{-\infty}^{\infty} E_\alpha(i^\alpha \omega^\alpha t^\alpha) f_\omega^{F,\alpha}(\omega)(d\omega)^\alpha,$$

then we can write
$$f_k(k\Delta t) \approx \frac{1}{(2\pi)^\alpha} \sum_{n=0}^{N-1} f_\omega^{F,\alpha}(n\Delta\omega)(\Delta\omega)^\alpha E_\alpha\left(i^\alpha t^\alpha n^\alpha (\Delta\omega)^\alpha\right) \quad (2.15)$$

with $\omega = N\Delta\omega$.

Taking $t = k\Delta t$ and $\dfrac{2\pi}{T} = \Delta\omega$ in (2.15) implies that
$$\varphi(k)$$
$$= f_k(k\Delta t)$$
$$\approx \frac{1}{(2\pi)^\alpha} \sum_{n=0}^{N-1} f_\omega^{F,\alpha}(n\Delta\omega)(\Delta\omega)^\alpha E_\alpha\left(i^\alpha n^\alpha (\Delta t)^\alpha k^\alpha (\Delta\omega)^\alpha\right)$$
$$= \frac{1}{T^\alpha} \sum_{n=0}^{N-1} \phi(n) E_\alpha\left(i^\alpha n^\alpha k^\alpha (2\pi)^\alpha / N^\alpha\right). \quad (2.16)$$

Combing the formulas (2.14) and (2.16), we have the following results:
$$\phi(n) = \frac{1}{\Gamma(1+\alpha)} \frac{T^\alpha}{N^\alpha} \sum_{k=0}^{N-1} \varphi(k) E_\alpha\left(-i^\alpha (2\pi)^\alpha n^\alpha k^\alpha / N^\alpha\right) \quad (2.17)$$

and
$$\varphi(k) = \frac{1}{T^\alpha} \sum_{n=0}^{N-1} \phi(n) E_\alpha\left(i^\alpha n^\alpha k^\alpha (2\pi)^\alpha / N^\alpha\right). \quad (2.18)$$

Setting $F(n) = \dfrac{1}{T^\alpha}\phi(n)$ and interchanging $k$ and $n$, we get
$$\varphi(n) = \sum_{k=0}^{N-1} F(k) E_\alpha\left(i^\alpha n^\alpha k^\alpha (2\pi)^\alpha / N^\alpha\right) \quad (2.19)$$

and
$$F(k) = \frac{1}{\Gamma(1+\alpha)} \frac{1}{N^\alpha} \sum_{n=0}^{N-1} \varphi(n) E_\alpha\left(-i^\alpha (2\pi)^\alpha n^\alpha k^\alpha / N^\alpha\right). \quad (2.20)$$

# 3 Discrete Yang-Fourier transforms of discrete-time fractal signal

*Definition 1*

Suppose that $F(k)$ be a periodic discrete-time fractal signal with period $N$. From (2.20) the sequence $f(n)$ is defined by

$$F(k) = \frac{1}{\Gamma(1+\alpha)} \frac{1}{N^\alpha} \sum_{n=0}^{N-1} f(n) E_\alpha \left( -i^\alpha (2\pi)^\alpha n^\alpha k^\alpha / N^\alpha \right), \qquad (3.1)$$

which is called $N$-point discrete Yang-Fourier transform of $F(n)$, denoted by

$$f(n) \leftrightarrow F(k).$$

*Definition 2*

Inverse discrete Yang-Fourier transform
From (2.19), the transform assigning the signal $F(k)$ to $f(n)$ is called the inverse discrete Yang-Fourier transform, which is rewritten as

$$f(n) = \sum_{k=0}^{N-1} F(k) E_\alpha \left( i^\alpha n^\alpha k^\alpha (2\pi)^\alpha / N^\alpha \right) \qquad (3.2)$$

Suppose that $f(n) \leftrightarrow F(k)$, $f_1(n) \leftrightarrow F_1(k)$ and $f_2(n) \leftrightarrow F_2(k)$, the following relations are valid:

*Property 1*

$$af_1(n) + bf_2(n) \leftrightarrow aF_1(k) + bF_2(k). \qquad (3.3)$$

Proof. Taking into account the linear transform of discrete Yang-Fourier transform, we directly deduce the result.

*Property 2*

Let $f(k)$ be a periodic discrete fractal signal with period $N$. Then we have

$$\sum_{n=j}^{j+N-1} f(n) = \sum_{n=0}^{N-1} f(n). \qquad (3.4)$$

Proof. We directly deduce the result when $j = mN + l$ with $0 \leq l \leq N-1$.

*Theorem 3*

Suppose that

$$F(n) = \frac{1}{\Gamma(1+\alpha)} \frac{1}{N^\alpha} \sum_{k=0}^{N-1} f(k) E_\alpha \left( -i^\alpha (2\pi)^\alpha n^\alpha k^\alpha / N^\alpha \right),$$

then we have

$$f(k) = \sum_{n=0}^{N-1} F(n) E_\alpha \left( i^\alpha n^\alpha k^\alpha (2\pi)^\alpha / N^\alpha \right) \qquad (3.5)$$

*Proof.* From the formulas (2.11)-(2.20) we deduce to the results.

# 4 Conclusions

In the present paper we discuss a model for the Yang-Fourier transforms of discrete approximation. As well, we give the discrete Yang-Fourier transforms of fractal signal as follows:

$$F(k) = \frac{1}{\Gamma(1+\alpha)} \frac{1}{N^\alpha} \sum_{n=0}^{N-1} f(n) E_\alpha \left( -i^\alpha (2\pi)^\alpha n^\alpha k^\alpha / N^\alpha \right)$$

and

$$f(n) = \sum_{k=0}^{N-1} F(k) E_\alpha \left( i^\alpha n^\alpha k^\alpha (2\pi)^\alpha / N^\alpha \right).$$

Furthermore, some results are discussed.